\documentclass[twocolumn,twoside,slac_two]{revtex4}
\usepackage{graphicx}
\usepackage{fancyhdr}
\def \b{{\cal B}}

\def \bo{B^0}

\def \ob{\overline{B}^0}

\def \s{\sqrt{2}}

\newcommand{\bi}{\begin{itemize}}
\newcommand{\ei}{\end{itemize}}
\newcommand{\bed}{\begin{description}}
\newcommand{\eed}{\end{description}}
\newcommand{\ben}{\begin{enumerate}}
\newcommand{\een}{\end{enumerate}}
\newenvironment{Eqnarray}{\arraycolsep
0.14em\begin{eqnarray}}{\end{eqnarray}}

\def\bea{\begin{Eqnarray}}
\def\eea{\end{Eqnarray}}
\def\beq{\begin{equation}}
\def\eeq{\end{equation}}

\begin{document}

\title{Direct CP Violation in $B$ Decays}

%

\author{M. Gronau}
\affiliation{Physics Department, Technion - Israel Insitute of Technology, 32000 Haifa, Israel}

\begin{abstract}
We discuss several aspects of direct CP asymmetries in $B$ decays, which are very useful in spite 
of hadronic uncertainties in asymmetry calculations. 1) Asymmetries in decays to 
$D^{(*)}K^{(*)}$, $\pi^+\pi^-,~\rho^+\rho^-$, providing precision tests for the CKM phase $\gamma$.
2) Null tests in $B^+\to J/\psi K^+, \pi^+\pi^0$, where a nonzero asymmetry provides 
evidence for New Physics. 3) Isospin and  
broken flavor SU(3) relations among CP asymmetries in $B\to K\pi,~\pi\pi$ predicting 
$A_{CP}(B^0\to K^0\pi^0)$ and $A_{CP}(B^0\to \pi^0\pi^0)$.
4) The significance of $A_{CP}(B^0\to K^+\pi^-)\ne A_{CP}(B^+\to K^+\pi^0)$. 
5) A potentially stringent constraint on $\gamma$ from $A_{CP}(B^+\to K^+\pi^0)$ and 
$R_c\equiv 2\Gamma(B^+\to K^+\pi^0)/\Gamma(B^+\to K^0\pi^+)$.
6) The role of direct CP asymmetries in $b\to s\bar qq$ decays for studying the origin of 
potential New Physics. 
\end{abstract}

\maketitle

\thispagestyle{fancy}


\section{Importance of direct CP violation}
It took 35 years from the discovery of tiny CP violation in $K^0$-$\bar K^0$ 
mixing by Christenson, Cronin, Fitch and Turlay~\cite{Christenson:1964fg} 
to an observation of direct CP violation in 
$K\to \pi\pi$ by the KTeV~\cite{Alavi-Harati:1999xp} and NA48~\cite{Fanti:1999nm}
collaborations. While this observation 
was very important by itself, ruling out the Superweak hypothesis for CP 
violation~\cite{Wolfenstein:1964ks}, hadronic uncertainties involved in calculating this effect 
prohibited a precise quantitative test of the CKM framework~\cite{Pich:2000hq}.

Tremendous effort has been devoted by the CLEO collaboration at Cornell,  by BaBar at SLAC,
Belle at KEK, and by the CDF and D0 collabotaions at Fermilab, to measure direct CP violation in hundreds of charged and neutral $B$ decay modes. A small sample of the measured 
asymmetries is plotted in Fig. 1~\cite{hfag}. The asymmetry in $B^0\to K^+\pi^-$, involving 
the smallest experimental error, provides unambiguous evidence for direct CP violation.
\begin{figure}[h]
\centering 
\includegraphics[width=80mm]{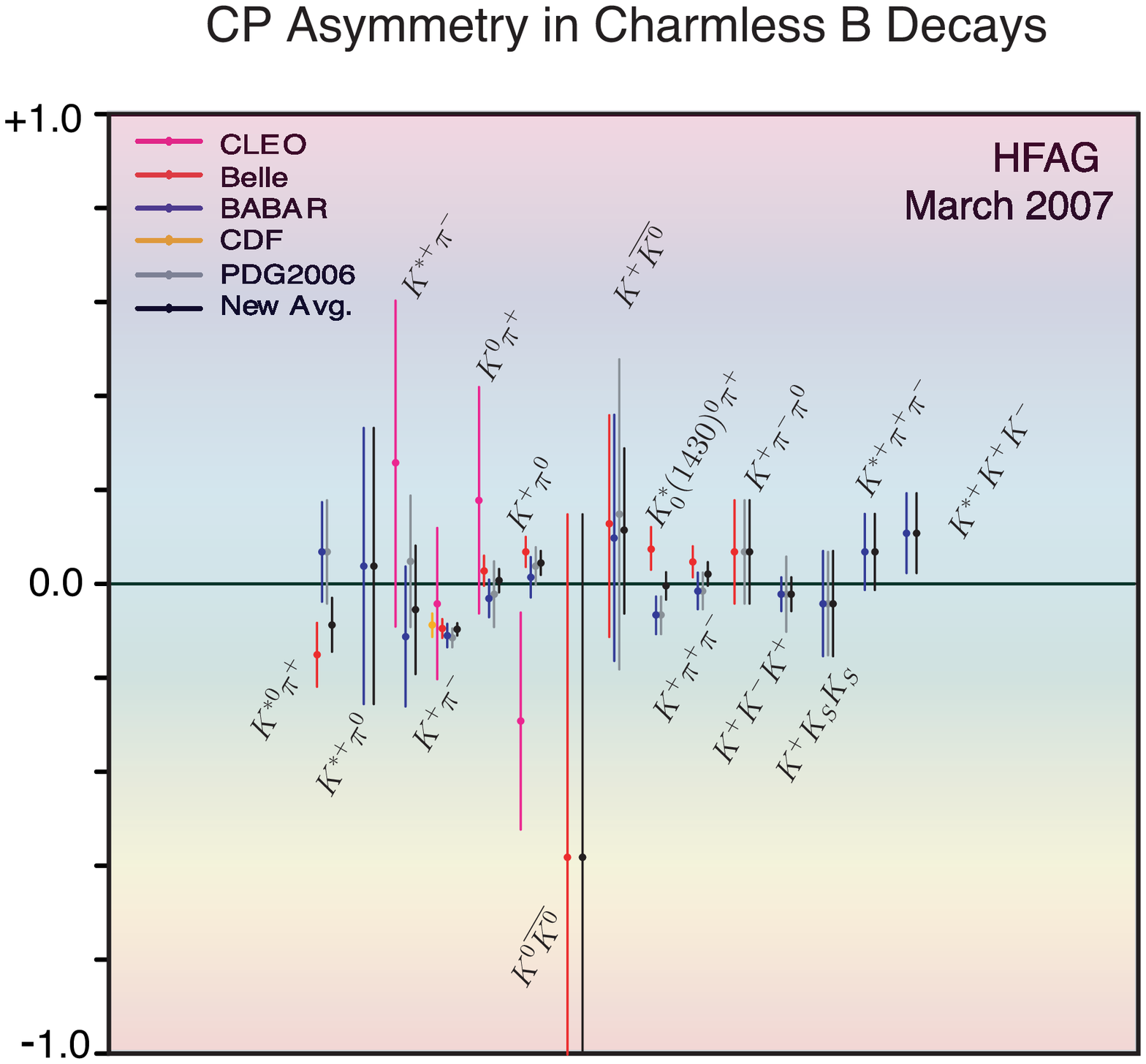} 
\caption{A sample of direct CP asymmetries~\cite{hfag}.} 
\label{fig:asym} 
\end{figure} 

\subsection{Difficulty of calculating asymmetries}
Calculations of direct CP asymmetries 
involve uncertainties from weak hadronic matrix elements and strong final state phases. 
To illustrate these uncertainties, consider for instance the decay $B^0\to K^+\pi^-$ which 
has a dominant penguin amplitude and a CKM-suppressed tree amplitude, as shown in Fig.~2.
\begin{figure}[h]
\includegraphics[width=80mm]{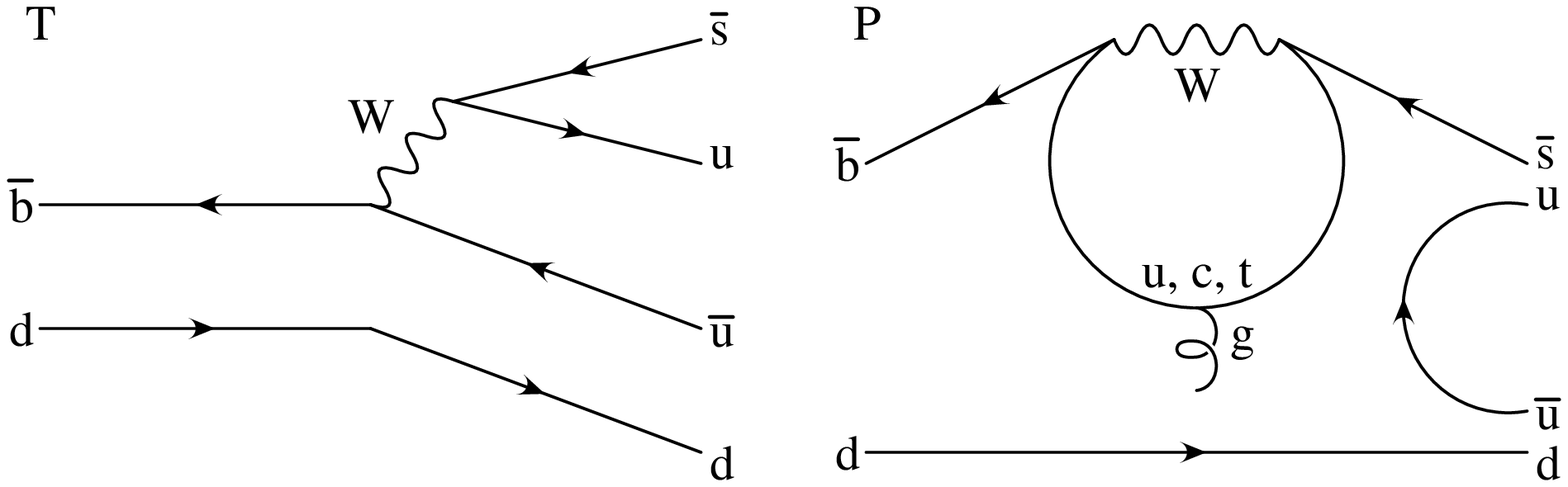} 
\caption{Penguin and tree amplitudes in $B^0\to K^+\pi^-$.} 
\label{treepen_s}
\end{figure}  
The amplitudes for this decay process and its charge-conjugate are given in terms of 
suitably defined magnitudes $|P|, |T|$, a CP-conserving strong phase $\delta$, and a
CP-violating weak phase $\gamma\equiv {\rm arg}(-V^*_{ub}V_{ud}/V^*_{cb}V_{cd})$,
\bea
A(\bo\to K^+\pi^-) & = & |P|e^{i\delta} + |T|e^{i\gamma}~,
\nonumber\\
A(\ob \to K^-\pi^+) & = & |P|e^{i\delta} + |T|e^{-i\gamma}~.
\eea
A calculation of the CP asymmetry in terms of $\gamma$,
\bea
A_{CP}(K^+\pi^-)  & \equiv &  
\frac{\Gamma(\ob\to K^-\pi^+) - \Gamma(\bo\to K^+\pi^-)}
{\Gamma(\ob\to K^-\pi^+) + \Gamma(\bo\to K^+\pi^-)}
\nonumber\\
& = &  -\frac{2|T/P|\sin\delta\sin\gamma}
{1 + |T/P|^2 + 2|T/P|\cos\delta\cos\gamma},
\eea
requires computing $|T/P|$ and $\delta$. This is extremely difficult, as these quantities involve 
non-perturbative long-distance effects. In QCD calculations based on a heavy quark 
expansion~\cite{Beneke:1999br,pQCD,SCET} one faces uncertainties in these quantities  
from chirally enhanced $1/m_b$-suppressed terms including
annihilation contributions from penguin operators, $\alpha_s$-suppressed terms  
and ``charming penguin" terms~\cite{Ciuchini:1997hb}. Some of these contributions can be 
traced back to incalculable soft rescattering amplitudes from $(\bar sc)(\bar c d)$ intermediate states.
A clear distinction between calculable short distance contributions and incalculable soft contributions 
is particularly challenging for the strong phase $\delta$.

While observing direct CP violation in $B$ decays is important by itself, it then seems that these asymmetries (like the one in $K\to\pi\pi$) cannot 
provide accurate tests for the mechanism of CP violation, originating in the Standard Model in the 
phase $\gamma$ of the Cabibbo-Kobayashi-Maskawa (CKM) matrix. {\em The purpose 
of this talk is to show that, in fact, direct asymmetries measured in certain $B$ decay modes 
do provide precision tests for the CKM framework, in spite of theoretical difficulties in calculating 
these asymmetries.} 

\subsection{Determining $\gamma$ in $B\to D^{(*)}K^{(*)}$}
We recall well-known examples of direct CP asymmetries in a whole class of processes 
$B\to D^{(*)}K^{(*)}$, where $D^0$ and $\bar D^0$ decay to a variety of common final states. These 
decays provide a clean determination of the weak phase $\gamma$~\cite{Gronau:1990ra}. 
The trick here lies in recognizing the measurements which yield this fundamental CP-violating 
quantity through interference of tree-level $\bar b\to \bar c u \bar s$ and $\bar b\to \bar u c \bar s$ amplitudes. A broad and up-to-date review of CP violation in the $B$ meson system, 
including numerous references for studies of $B\to D^{(*)}K^{(*)}$, can be found in 
Ref.~\cite{Gronau:2007xg}. 

\subsection{Determining $\gamma$ in $B\to \pi^+\pi^-,~\rho^+\rho^-$}
In $B^0\to\pi^+\pi^-$ and $B^0\to\rho^+\rho^-$, mixing-induced asymmetries ($S$) {\em and
direct asymmetries} ($C\equiv -A_{CP}$) are both needed to fix $\gamma$ in a rather
precise method based on isospin symmetry~\cite{Gronau:1990ka}. 
Measurements of these asymmetries and conservative assumptions about flavor SU(3) breaking
yield the currently most precise value 
$\gamma =(72\pm 6)^\circ$~\cite{Beneke:2006rb,Gronau:2007af,Gronau:2007xg}, in 
agreement with $\gamma=(66\pm 6)^\circ$
obtained from $\Delta m_d/\Delta m_s$~\cite{Rosner07}. 

\subsection{Null tests of the CKM framework}
Interesting applications of direct CP asymmetries are null tests of the CKM framework
in decays where asymmetries are expected to be very small. Two processes
dominated by a single CKM phase are $B^+\to J/\psi K^+$~\cite{Gronau:1989ia,BMR}
and $B^+\to \pi^+\pi^0$, where the Standard Model predicts vanishingly small 
asymmetries, much smaller than one percent. This includes electroweak penguin 
contributions in $B^+\to\pi^+\pi^0$~\cite{Buras:1998rb}. A nonzero asymmetry observed
in one of these modes at a percent level (as small as it can be with future anticipated 
precision~\cite{hfag}) would be a clean signature for New Physics.

\subsection{Enhanced CP asymmetries}
While in general calculating strong phases is very difficult 
(though phases are known in e.g. $B^+\to K^+K^-K^+$ mediated by $c\bar c$ 
resonant states~\cite{Eilam:1995nz}),
asymmetry measurements provide information
for studying the dynamics of hadronic decays. Thus, certain charmless $B$ decays have been
predicted to lead to larger asymmetries than others, because they involve two interfering 
amplitudes with different weak phases, whose ratio is expected to be dynamically enhanced. 
This may follow from an enhancement of the smaller of the two amplitudes, or a suppression 
of the larger amplitude. Such effects were noted in an approach using flavor 
SU(3) symmetry ~\cite{Gronau:1994rj,Chiang:2004nm}, and may probably be realized in QCD 
calculations~\cite{Beneke:1999br,pQCD}.
We mention four examples for this enhancement effect using the language of flavor SU(3) 
amplitudes: (1) $B^+\to \pi^+\eta$, where an amplitude $P$ smaller than $T$ 
involves a factor 2, (2) $B^+\to K^+\eta$, where a potentially dominant $P$ amplitude involves
destructive interference between a few quark-level penguin amplitudes~\cite{Lipkin:1990us}, 
(3) $B^+\to K^+\rho^0$, where subdominant $T_V$ and $C_P$ amplitudes
add up constructively, (4) $B^0\to \rho^{\pm}\pi^{\mp}$, which involves constructive 
interference of $P_VT_V$ and $P_PT_P$ terms.

\begin{table}[h]
\caption{CP asymmetries involving large $|\Delta S|=1$ tree-to-penguin ratios or 
large $\Delta S=0$ penguin-to-tree ratios~\cite{hfag}.}
\label{tab:3sigma}
\begin{center}
\begin{tabular}{c c c c c} \hline
$\pi^+\eta~~~~~~~~~K^+\eta$&$K^+\rho^0$&$\rho^{\pm}\pi^{\mp}$
\\ \hline
$-0.19\pm 0.07-0.29\pm 0.11$&$0.31^{+0.11}_{-0.10}$&$-0.13\pm 0.04$
\\ \hline
\end{tabular}
\end{center}
\end{table}

Table 1 quotes measured CP asymmetries for these four final states~\cite{hfag}. All four 
asymmetries are nonzero at a level of about $3\sigma$ with central values between 13 
and 31 percent. Somewhat higher precision is required in these asymmetry measurements
for claiming unambiguous evidence of direct CP violation in these modes.

\section{CP asymmetries in $B\to K\pi$}
\subsection{Isospin sum rules for $\Gamma$ and $A_{CP}$}
The decays $B^{0,+}\to K\pi$, which occur in four distinct final states, can be expressed
in terms of three isospin amplitudes~\cite{Nir:1991cu}. 
The initial states are $I(B)=1/2$, the final states are $I(K\pi)=1/2,3/2$ and the effective
weak Hamiltonian consists of $\Delta I=0,1$. Denoting  
$\Delta I=0$ and $\Delta I=1$ amplitudes by $B$ and $A $ or $A'$, 
the physical amplitudes for the two pairs of $B^+$ and $B^0$ decays are related to 
each other by isospin reflection, implying~\cite{Gronau:2007ut}
\bea
&&A(K^0\pi^+) =  B+A,~-A(K^+\pi^-)=B-A,
\\
&&-\s A(K^+\pi^0) = B+A',~\s A(K^0\pi^0)=B-A'.
\nonumber
\eea
This leads to an isospin quadrangle relation,
\beq
\label{quadrangle}
A(K^0\pi^+)-A(K^+\pi^-)+\s A(K^+\pi^0)-\s A(K^0\pi^0)=0,
\eeq
which has important physical implications in terms of two approximate sum rules for decay 
rates $\Gamma$~\cite{Gronau:1998ep} and CP rate differences $\Delta$~\cite{Atwood:1997iw}:
\bea\label{SRKpi}
\Gamma(K^+\pi^-)+\Gamma(K^0\pi^+) &\approx &2[\Gamma(K^+\pi^0)+\Gamma(K^0\pi^0)],
\nonumber\\
\Delta(K^+\pi^-)+\Delta(K^0\pi^+) &\approx &2[\Delta(K^+\pi^0)+\Delta(K^0\pi^0)],
\nonumber\\
\eea
where
\beq\label{Delta}
\Delta(K\pi) \equiv \Gamma(\bar B\to \bar K\bar\pi)-\Gamma(B\to K\pi).
\eeq

We now present the shortest proofs for these sum rules, using the dominance of a 
$\Delta I=0$ penguin amplitude $P$ (part of the isospin amplitude $B$) in all $B\to K\pi$ 
decays. Evidence for penguin-dominance is provided by the four measured $B\to K\pi$ decay 
rates which are equal within $2\sigma$~\cite{hfag},
\bea\label{R}
R & \equiv & \frac{\Gamma(B^0 \to K^+ \pi^-)}
{\Gamma(B^+ \to K^0 \pi^+)} = 0.90 \pm 0.05,\cr
R_c  & \equiv  & \frac{2\Gamma(B^+ \to K^+ \pi^0)}
{\Gamma(B^+ \to K^0 \pi^+)} = 1.11 \pm 0.07,\cr
R_n  & \equiv & \frac{\Gamma(B^0 \to K^+ \pi^-)}
{2\Gamma(B^0 \to K^0 \pi^0)} = 0.97 \pm 0.07.
\eea
The non-penguin amplitudes are calculated to be much smaller than $P$, 
$|$non-$P|/|P|\sim 0.1$~\cite{Beneke:1999br,pQCD,SCET,Ciuchini:1997hb,Chiang:2004nm}. 
Terms in the two sum rules (\ref{SRKpi}) which are quadratic in $P$ (or $B$) cancel trivially,
while terms which are linear in $P$ cancel because of (\ref{quadrangle}). Thus, the only 
terms which may violate the two sum rules are quadratic in non-penguin amplitudes, and 
can be shown to amount to a few percent of each side of the two sum rules (\ref{SRKpi}).  

Indeed, the sum rule for decay rates $\Gamma$ holds within experimental errors which are 
currently about $5\%$ of each side~\cite{hfag,Gronau:2006xu}.
The sum rule for CP rate asymmetries $\Delta$, expected to hold within a similar precision, 
leads to a {\em prediction for the asymmetry in $B^0\to K^0\pi^0$} in terms of the other 
three asymmetries which have been measured with higher precision (see Table II),
\beq\label{ACPK0pi0}
A_{CP}(B^0\to K^0\pi^0) = -0.140\pm 0.043.
\eeq
This prediction, which is expected to hold within a few percent,
can be improved by reducing errors in $A_{CP}(K^0\pi^+), A_{CP}(K^+\pi^0)$.
While the value (\ref{ACPK0pi0}) is consistent with experiment,  higher accuracy in 
asymmetry measurements is required for testing this prediction following from 
the rather precise $\Delta $ relation (\ref{SRKpi}).
\begin{table}[h]
\caption{CP asymmetries in $B\to K\pi$~\cite{hfag}.}
\label{tab:AcpKpi}
\begin{center}
\begin{tabular}{c c c c c} \hline
$B^0\to K^+\pi^-$ & $B^+\to K^+\pi^0$ & $B^+\to K^0\pi^+$ & $B^0\to K^0\pi^0$\\ \hline
$-0.097\pm 0.012$ & $0.047\pm 0.026$ & $0.009\pm 0.025$ & $-0.12\pm 0.11$\\
\hline
\end{tabular}
\end{center}
\end{table}

\subsection{$A_{CP}(K^+\pi^0)\ne A_{CP}(K^+\pi^-)$ a puzzle?}
The measurement of a nonzero CP asymmetry in $B^0\to K^+\pi^-$ provides the  first evidence
for an interference between a dominant penguin amplitude $P$ and a small tree amplitude 
$T$ with a nonzero relative strong phase $\delta$ between the two amplitudes. [See Eqs.~(1) and (2)].
Such an interference occurs also in $B^+\to K^+\pi^0$, in which a spectator $d$-quark in $B^0\to K^+\pi^-$ is replaced by a $u$-quark.
No asymmetry has been observed in $B^+\to K^+\pi^0$. An assumption that other contributions to 
this asymmetry are negligible has raised some questions 
about the validity of the CKM framework. In fact,  a color-suppressed  tree amplitude $C$, 
also occurring in $B^+\to K^+\pi^0$~\cite{Gronau:1994rj} (see Fig.~3), resolves this ``puzzle" 
if this amplitude is comparable in magnitude to $T$. 
\begin{figure}[h]
\centering 
\includegraphics[width=30mm]{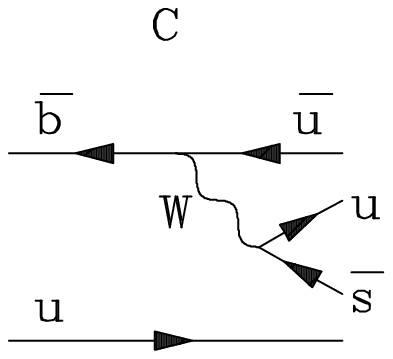} 
\caption{Color-suppressed tree amplitude in $B^+\to K^+\pi^0$} 
\label{fig:asym} 
\end{figure} 
A too naive assumption, $|C|\ll|T|$, has been made
in Ref.~\cite{Gronau:1994rj} followed by numerous other works.

More recent studies, including a global SU(3) fit for 
all charmless $B$ decays to two pseudoscalars,
have shown that $|C|\sim |T|$~\cite{Chiang:2004nm,Buras:2003dj,Baek:2004rp}.
For consistency between the two CP asymmetries in $B^0\to K^+\pi^-$ and 
$B^+\to K^+\pi^0$, the strong phase difference between $C$ and $T$ must be 
negative and cannot be very small~\cite{Gronau:2006ha}. This, and the somewhat large 
value of $C$, seem to stand in 
contrast to QCD calculations using a factorization theorem~\cite{Beneke:1999br,SCET}.
While this may be considered a difficulty for QCD calculations, by no means should it
be considered evidence for New Physics as argued sometimes.

\subsection{A sum rule for $A_{CP}(K^+\pi^0)$ and $R_c$}
The asymmetry $A_{CP}(K^+\pi^0)$ and the ratio of rates $R_c$ defined in (\ref{R})
involve the decay amplitude for $B^+\to K^+\pi^0$, 
which seems to confront QCD calculation with a difficulty. The smallness of the measured 
asymmetry and of the measured value of $R_c-1$ lead  to an interesting constraint on
$\gamma$ which we discuss now. 

Including color-favored and color-suppressed electroweak penguin 
contributions, $P_{EW}$ and $P^c_{EW}$, one has 
\bea
A(K^+\pi^0) & = &P+T+C+P_{EW}+P^c_{EW},
\nonumber\\
A(K^0\pi^+) &= &P.
\label{A+}
\eea
A small $1/m_b$-suppressed annihilation amplitude $A$ from a current-current 
operator has been neglected 
in the two processes~\cite{Beneke:1999br,SCET,Gronau:1994rj,Blok:1997yj}.
One introduces two calculable parameters for ratios of amplitudes,
$r_c\equiv |T+C|/|P|$ and $\delta_{EW}\equiv |P_{EW}+P^c_{EW}|/|T+C|$.
The parameter $r_c$ is given by~\cite{Gronau:1994bn} 
\beq 
r_c = \sqrt{2}\frac{V_{us}}{V_{ud}}\frac{f_K}{f_\pi}
\sqrt{\frac{\b(B^+\to\pi^+\pi^0)}
{\b(B^+\to K^0\pi^+)}} = 0.20 \pm 0.02,
\eeq
where the error includes an uncertainty from SU(3) breaking.
The parameter $\delta_{EW}$ is defined by~\cite{Neubert:1998pt}
\bea
P_{EW} &+& P^c_{EW} =  -\delta_{EW}e^{-i\gamma}(T+ C),
\nonumber\\
\delta_{EW} & = & -\frac{3}{2}\frac{c_9 + c_{10}}{c_1 + c_2}
\frac{|V^*_{tb}V_{ts}|}{|V^*_{ub}V_{us}|} = 0.60 \pm 0.05.
\eea
Here the error is dominated by the current uncertainty in 
$|V_{ub}|/|V_{cb}|$, including also a 
smaller error from SU(3) breaking estimated using QCD factorization.

The asymmetry $A_{CP}(K^+\pi^0)$ and the ratio of rates $R_c$ are given, 
to first order in $r_c$, by
\bea
A_{CP}(K^+\pi^0)&=&-2r_c\sin\gamma\sin\delta_c +{\cal O}(r^2_c),
\\
R_c - 1&=& -2r_c(\cos\gamma-\delta_{EW})\cos\delta_c +{\cal O}(r^2_c),
\nonumber
\eea 
where $\delta_c$ is the strong phase difference  between $T+C$ and $P$.
Eliminating $\delta_c$
it is now straight forward to prove the following sum rule~\cite{Gronau:2006ha}
\beq\label{SRK+pi0}
\left( \frac{A_{CP}(K^+ \pi^0)}{\sin \gamma} \right)^2 
+ \left( \frac{R_c-1}{\cos \gamma - \delta_{\rm EW}} \right)^2 
= (2r_c)^2 + {\cal O}(r_c^3).
\eeq

This sum rule implies that at least one of the two terms whose
squares occur on the left-hand-side must be sizable, of the order of
$2r_c=0.4$. The first term, $|A_{CP}(B^+\to K^+\pi ^0)|/\sin\gamma$, is
already smaller than $\simeq 0.1$, using the current $2\sigma$ bounds on $\gamma$ and  
$|A_{CP}(B^+\to K^+\pi^0)|$. Thus, the second term must provide a
dominant contribution. For $R_c\simeq 1$, this implies $\gamma\simeq 
\arccos\delta_{EW} \simeq (53.1\pm 3.5)^\circ$. This range is expanded by including 
errors in $R_c$ and $A_{CP}(B^+\to K^+\pi^0)$. 
For instance, bounds $0.95 < R_c <  1.1$ would imply an important upper limit, 
$\gamma < 71^\circ$. Current values of $A_{CP}(K^+\pi^0)$ and $R_c$ lead to 
an upper limit $\gamma \le 88^\circ$ at 
$90\%$ confidence level~\cite{Gronau:2006ha}. This bound is consistent with 
the value of $\gamma$ obtained from $B\to\pi^+\pi^-$ and $B\to\rho^+\rho^-$, 
as mentioned above, but is not yet competitive with its precision. 

\section{SU(3) relations $\Delta(K\pi)=-\Delta(\pi\pi)$}
One may prove two useful relations between CP rate differences within the CKM
framework~\cite{Deshpande:1994ii,Gronau:1995qd}: 
\bea\label{+-}
\Delta(K^+\pi^-) & = & -\Delta(\pi^+\pi^-),\\
\label{00}
\Delta(K^0\pi^0) & = & -\Delta(\pi^0\pi^0).
\eea
A slightly over-simplified proof of these relations goes as follows. [A precise 
proof, including electroweak penguin terms and justifying an assumption about 
negligible $E+PA$ terms can be found in Ref.~\cite{Gronau:2007xg}).]
Writing 
\beq
A(K^+\pi^-) = P + T,
\eeq
where $P$ and $T$ contain strong and weak phases,
one has in the flavor SU(3) limit
\beq
A(\pi^+\pi^-) = -\lambda P +\lambda^{-1}T,
\eeq
where $\lambda \equiv V_{us}/V_{ud}=-V_{cd}/V_{cs}$. Similarly,
\bea
\s A(K^0\pi^0) & = & P-C,
\nonumber\\
\s A(\pi^0\pi^0) & = & -\lambda P -\lambda^{-1}C.
\eea
The CP rate differences in the two pairs of processes are given by interference terms
between $P$ and $T$ and between $P$ and $C$, which are equal in magnitude and 
have opposite signs in $B\to K\pi$ and $B\to\pi\pi$. This proves (\ref{+-}) and (\ref{00}).

\begin{table}[h]
\caption{Direct CP asymmetries in $B\to \pi\pi$~\cite{hfag}.}
\label{tab:Acppipi}
\begin{center}
\begin{tabular}{c c} \hline
$B^0\to \pi^+\pi^-$ & $B^0\to \pi^0\pi^0$\\ \hline
$0.38\pm 0.07$ & $0.36^{+0.33}_{-0.31}$\\
\hline
\end{tabular}
\end{center}
\end{table}
Using branching ratios from~\cite{hfag} and asymmetries in Tables II and III, 
Eq.~(\ref{+-}) reads
\bea\label{+-exp}
\b(K^+\pi^-)A_{CP}(K^+\pi^-) &=& - \b(\pi^+\pi^-)A_{CP}(\pi^+\pi^-),
\nonumber\\
(-1.88\pm 0.24)10^{-6}&=& (-1.96\pm 0.37)10^{-6}.
\eea
Both the signs and the magnitudes agree well, providing evidence for the success of 
flavor SU(3).
We note that using SU(3) breaking factors $f_K/f_\pi$ for both $T$ and $P$, 
as assumed in~\cite{Deshpande:1994ii}, would imply a factor 
$(f_K/f_\pi)^2$ on the right-hand-side of (\ref{+-exp}) leading to a worse agreement.
Some reduction of errors in $A_{CP}(K^+\pi^-)$ and $A_{CP}(\pi^+\pi^-)$ is required in
order to determine well the pattern of SU(3) breaking in $PT$.

The relation (\ref{00}) and the value of $A_{CP}(K^0\pi^0)$ in (\ref{ACPK0pi0}) obtained
using isospin symmetry imply a {\em prediction for $A_{CP}(\pi^0\pi^0)$},
\beq
A_{CP}(\pi^0\pi^0)=-A_{CP}(K^0\pi^0)\frac{\b(K^0\pi^0)}{\b(\pi^0\pi^0)}
=1.07\pm 0.38.
\eeq
The error is dominated by errors in $A_{CP}(K^0\pi^+)$ and $A_{CP}(K^+\pi^0)$.
An SU(3) breaking factor $f_K/f_\pi$ in $C$ would lower this prediction by a factor $f_\pi/f_K$.
A large positive CP asymmetry in $B^0\to\pi^0\pi^0$ implies comparable sides in the
$\bar B\to\pi\pi$ isospin amplitude triangle, but a squashed $B\to\pi\pi$ isospin  triangle. 
This has a simplifying effect on the isospin analysis in $B\to\pi\pi$, where a discrete ambiguity 
disappears in the limit of a flat $B\to\pi\pi$ triangle~\cite{Gronau:1990ka}.

\section{CP violation in $b\to s\bar qq$}
A class of $b\to s\bar qq$ penguin-dominated $B^0$ decays to CP-eigenstates has  
attracted recently considerable attention.  
This includes 
final states $XK_S$ and $XK_L$, where $X=\phi,  \pi^0, \eta',  \omega, f_0, \rho^0, K^+K^-, K_SK_S, 
\pi^0\pi^0$, for which measured mixing-induced asymmetries $\pm S$ (for CP eignstates
with eigenvalues $\eta_{CP}=\mp 1$) and {\em direct asymmetries} $C\equiv -A_{CP}$ are quoted 
in Table IV~\cite{hfag}. 
\begin{table}[h]
\caption{CP asymmetries in $b\to s\bar qq$ for $\eta_{CP}=\mp 1$~\cite{hfag}.}
{\begin{tabular}{@{}cccccc@{}} 
\hline\hline
$X$&$\phi$&$\pi^0$&$\eta'$&$\omega$\\
$\pm S$&$0.39\pm 0.18$&$0.33\pm 0.21$&$0.61\pm 0.07$&
$0.48\pm 0.24$\\
$C$&$0.01\pm 0.13$&$0.12\pm 0.11$&$-0.09\pm 0.06$&
$-0.21\pm 0.19$\\
\hline
$X$&$\rho^0$&$f_0(980)$&$K^+K^-$&$K_SK_S$\\
$\pm S$&$0.20\pm 0.57$&$0.42\pm 0.17$&$0.58^{+0.18}_{-0.13}$&
$0.58\pm 0.20$\\
$C$&$0.64\pm 0.46$&$-0.02\pm 0.13$&$0.15\pm 0.09$&$-0.14\pm 0.15$\\
\hline\hline
\end{tabular}}
\end{table}

Whereas a value $S= -\eta_{CP}\sin2 \beta=0.678\pm 0.025$~\cite{hfag} is expected 
approximately~\cite{London:1989ph,Grossman:1996ke}, the actual average of all 
corresponding measured entries
in Table IV is $\sin\beta_{\rm eff} \equiv \langle -\eta_{CP}S\rangle = 0.53\pm 0.05$.
A question often raised is ``{\em is this $2.6\sigma$ discrepancy caused by 
New Physics?}". In a similar manner, one may calculate the {\em average of all direct 
CP asymmetries} obtaining a value $\langle A_{CP}\rangle = 0.01\pm 0.04$, which is 
small and consistent with zero.
{\em Does this mean that there is very little place for New Physics?}

These two questions, 
considering averages over several processes,
should be considered with care because in the Standard Model 
both $\Delta S\equiv S+\eta_{CP}\sin 2\beta$ and $C\equiv -A_{CP}$ are process-dependent. 
The smallness of the asymmetries $A_{CP}$ relative to $\Delta S$ may be related to small
strong phases $\delta$, because $A_{CP}$ and $\Delta S$ are proportional to $\sin\delta$ 
and $\cos\delta$, respectively~\cite{Gronau:1989ia}.
Calculations  of these asymmetries involve 
hadronic uncertainties at a level of several percent, of order 
$\lambda^2$~\cite{Grossman:2003qp,Beneke:2005pu,Williamson:2006hb,Cheng:2005bg}.
It has been pointed out some time ago~\cite{Atwood:1997zr} that if New Physics contributions 
are at this low level, it becomes difficult to separate them from  hadronic uncertainties 
within the CKM framework.

The importance of {\it direct CP asymmetries} measured in this class of processes may be 
demonstrated through two features of $C$ and $\Delta S$, by which one can  distinguish 
New Physics effects from hadronic uncertainties in the Standard Model and learn about
the origin of New Physics effects:
\begin{itemize}
\item Within the Standard Model $C$ and $\Delta S$ can be shown to lie on 
a circle~\cite{Gronau:1989ia},
\beq
\left(\frac{\Delta S}{\cos 2\beta}\right)^2 + C^2 = \left(2\xi\sin\gamma\right)^2,
\eeq
whose radius depends on a process-dependent ratio of tree and penguin amplitudes
$\xi\sim {\cal O}(\lambda^2)$. The locus on the circle is fixed by a strong phase
$\delta$. 
In most cases one expects $|\delta|<\pi/2$ implying 
$\Delta S>0$~\cite{Grossman:2003qp,Beneke:2005pu,Williamson:2006hb,Cheng:2005bg}.
\item Once the measured values of $C$ and $\Delta S$ disagree 
with calculations of $\xi$ and the strong phase $\delta$ beyond hadronic 
uncertainties, we will have solid evidence for New Physics. At this point one would 
seek signatures characterizing classes of models rather than studying effects of 
specific models of which quite a few exist
\cite{Grossman:1996ke,Gronau:1996rv,Ciuchini:1997zp,Hiller:2002ci,Deshpande:2003nx,
Barger:2003hg}.
A useful way for classifying extensions of the Standard Model 
is by the isospin behavior, $\Delta I=0$ or $\Delta I=1$, of the new effective operators.  
\end{itemize}

Recently it has been shown~\cite{Gronau:2007ut} that the isospin structure of potential New
Physics operators contributing to $b\to s\bar qq$ can be determined
 by studying $C$ and $\Delta S$ in $B^0\to XK^0$ together with two other kinds 
 of asymmetries: direct asymmetries $A_{CP}$ in isospin-reflected  decays $B^+\to XK^+$, 
 and isospin-dependent CP-conserving asymmetries defined by
\beq
A_I\equiv\frac{\Gamma(XK^+)-\Gamma(XK^0)}{\Gamma(XK^+)+\Gamma(XK^0)}.
\eeq
A study of the currently measured four kinds of asymmetries has shown that potential New Physics contributions to these processes must be small. Some reduction of errors in the measured asymmetries is required for identifying a signature for New Physics and for a useful implementation of  this method.
We refer the reader to Ref.~\cite{Gronau:2007ut} for details of this analysis.

\section{Conclusion}
The importance of direct CP violation is demonstrated by using direct asymmetries in 
$B\to D^{(*)}K^{(*)}, \pi^+\pi^-, \rho^+\rho^-$ for a determination of the weak phase 
$\gamma$. Asymmetries in $B^+\to J/\psi K^+, \pi^+\pi^0$ provide unambiguous 
signatures for New Physics. In spite of the difficuly of calculating strong phases, measured
asymmetries provide useful information about the dynamics of hadronic decays.

The different asymmetries measured in $B^+\to K^+\pi^0$ and $B^0\to K^+\pi^-$
cannot be easily explained within QCD calculations, but should not be considered evidence
for New Physics. An isospin sum rule among the four $B\to K\pi$ asymmetries 
predicts $A_{CP}(K^0\pi^0)=-0.140\pm 0.043$. Small $A_{CP}(K^+\pi^0)$ and small $R_c-1$
 imply an interesting constraint on $\gamma$. 
 The flavor SU(3) prediction $A_{CP}(K^+\pi^-)/A_{CP}(\pi^+\pi^-)=-\b(\pi^+\pi^-)/\b(K^+\pi^-)$
 works well. The ratio of the two asymmetries fixes the pattern
 of SU(3) breaking, which is useful for a precise determination of $\gamma$. Flavor SU(3) 
 predicts a large positive direct asymmetry in $B\to\pi^0\pi^0$, which has an implication on the $B\to\pi\pi$ isospin analysis. Direct CP asymmetries in $b\to s\bar q q$ play a central role in 
 studying  New Physics operators, in particular for learning their isospin structure.

\begin{acknowledgments}
I would like to thank the Local Organizing Committee for a very interesting and smoothly 
running conference in a beautiful setting.
I am grateful to Jonathan Rosner for a long and fruitful collaboration and to several other
short-term collaborators.
This work was supported in part by the Israel Science Foundation
under Grant No.\ 1052/04 and by the German-Israeli Foundation under
Grant No.\ I-781-55.14/2003.
\end{acknowledgments}

\end{document}